\magnification=1200

\overfullrule=0pt
\hoffset=3truecm\hsize=14truecm
\baselineskip=24pt
\vsize=24truecm\voffset=2truecm
\def\frac#1#2{{#1\over#2}}

{\bf Evolution of Gaussian Wave Packet and Nonadiabatic Geometrical Phase for the
time-dependent Singular Oscillator}

\centerline {M. Maamache and H. Bekkar}\smallskip \centerline
{Laboratoire de Physique Quantique et Systemes Dynamiques,} 
\centerline{D\'epartement de
Physique, Facult\'e des Sciences,}\centerline { Universit\'e de S\'etif, S\'etif (19000)
ALGERIA} \bigskip
\noindent  The geometrical phase of a time-dependent singular oscillator is
obtained in the framework of Gaussian wave packet. It is shown by a simple geometrical
approach that the geometrical phase is connected to the classical nonadiabatic Hannay
angle of the generalized Harmonic oscillator.

\bigskip\noindent PACS
number( s ): $03.65.Bz, 03.65.Ge, 03.65.Sq$

Explicitly time-dependent problems present special difficulties in classical
and quantum mechanics. However, they deserve detailed study because very interesting
properties emerge when, even for simple linear systems, some parameters are allowed to
vary with time. For instance, particular recent interest has been devoted to systems in
which evolution originates geometric contributions [1-6]. One of these, the generalized
harmonic oscillator has invoked much attention to study the nonadiabatic geometric phase
for various quantum states, such as Gaussian, number, squeezed or coherent states, which
can be found exactly [7-10]. Recently, the geometric phase for a cyclic
wave packet solution of the generalized harmonic oscillator and its relation to Hannay's
angle  were studied by Ge and Ghild [7]. They introduce the time-dependent Heller
Gaussian wave packet form [11] $$ \Phi (x, t) = {\hbox{exp}}\bigl(\hbar^{-1}\bigl[ -\alpha
(x - q)^{2} + ip(x -q) + k\bigr]\bigl)\eqno(1)$$ centred around the classical guiding
trajectory $(q,p)$,  and proceed to derive equations of motion for the complex or real
parameters ($\alpha(t),\ q(t),\ p(t),$ and $k(t)$) which serve to specify a complete
quantum wave packet. 

On the other hand, the number of exactly solvable quantum time-dependent problems is very
restricted,  one of the rare
examples admitting exact solutions of the Schr\"odinger equation and have been studied intensively
lately [12-19] is the quantum time-dependent generalized singular oscilator $$ H = {1\over
2}  \bigl\lbrack Z(t) p_x^{2}
+ Y(t)( p_x x + xp_x) + X(t) x^{2} + {{Z(t) l^{2}}\over {x^{2}}}\bigr\rbrack \eqno (2)$$
where $x$ and $p_x = - \imath\hbar {\partial / {\partial x}}$ are the quantum
operators, $X(t), Y(t)$, and $Z(t)$ are an arbitrary function of
time, and $l$ is an arbitrary constant which could be zero. A distinguished role of the
Hamiltonian (2) is explained by the fact that, in a sense, it belongs to a boundary
between linear and nonlinear problems of classical and quantum mechanics. For this reason,
it was used in many applications in different areas of physics. For example, it served as
an initial point in constructing interesting exactly solvable models of interacting N-body
systems [12-13]. It was also used for modeling diatomic and polyatomic molecules [14]. It
can have some relation to the problem of the relative motion of ions in electromagnetic
traps [19].

The aim of this letter is to explore Gaussian wave function dynamics for the Hamiltonian
(2) with nonadiabatic time dependence, and formulate a geometrical approach to derive
a nonadiabatic geometric phase effect in quantum and classical mechanics. For that
purpose we intoduce a class of wave function of the form $$\Psi_l(x,t) = x^{\bigl(1/2
-\sqrt{(l/\hbar)^2 + 1/4 }\bigr)}\, {\hbox{exp}}\Biggl\{{1\over\hbar}\Biggl(1/2 (l +
ipq)\Bigl[  ({{x - q}\over q})^{2} + 2 ({{x -q}\over q})  \Bigr] + k\Bigr)\Bigr\}\eqno
(3)$$ given as products of squeezed Gaussian wave packet of the type (1) and a function
$x$ of order $(1/2 -\sqrt{(l/\hbar)^2 + 1/4 })$. Inserting Eq. (3) into Schr\"odinger
equation $$i\hbar{{\partial\Psi_l}\over{\partial t}} = H\Psi_l ,\eqno (4)$$ and then
compare between the coefficients of various powers of $( x- q)$. This lead to $$(x-
q)^{2}:\ i\dot{\beta} = 2Z\beta^{2} - 2iY\beta + {X\over 2}, \eqno (5)$$ where $\beta =
-ip/(2 q) - l/(2 q^2),$ $$(x- q)^{1} :\ - 2i(\beta - { l\over {2 q^{2}}}) (\dot{q} - Zp -
Yq) + (\dot{p} + Xq + Yp - {{Z l^{2}}\over {q^{3}}}) = 0, \eqno (6)$$ $(x- q)^{0}$:
$$p\dot{q} +i {\dot k} = {1\over 2}  \bigl\lbrack Z p^{2} + 2Y p q  + X q^{2} + {{Z
l^{2}}\over {q^{2}}}\bigr\rbrack - {{Z(t) l^{2}}\over {q^{2}}} + 2\hbar (1/2
-\sqrt{(l/\hbar)^2 + 1/4 })(Z\beta -i{Y\over 2}).\eqno (7)$$ 

The $(x- q)^{2}$ condition determines $\beta$ by a nonlinear equation of the Riccati
form, which can be transformed to a linear system by introducing a two dimensional
vector $\vec{ v}^{T} \equiv (Q, P)$ and $$\beta \equiv -{i\over 2}{P\over Q},\eqno (8)$$
where $Q$ and $P$ may be complex. In order that $\beta$ satisfies (5), it is sufficient
that $\vec{v}$ obey the Hamilton's equation $$\vec{\dot v} = \pmatrix{\dot
{Q}\cr\dot{P}\cr}= \pmatrix{ {\ Y}&{\ Z}\cr{-X}&{-Y}\cr}\pmatrix{{Q}\cr{P}\cr} = -{\cal
H}\vec v\, .\eqno (9)$$ The $(x- q)^{1}$ condition makes sense if $$\eqalignno {\dot{q} &= Zp
+ Yq \cr \dot{p} &= - Xq - Yp + {{Z l^{2}}\over {q^{3}}}\ ,&(10)\cr}$$ and determines 
the complex guiding trajectory associated to the classical Hamiltonian $$H(q,p,t) = {1\over
2}  \bigl\lbrack Z(t) p^{2}
+ Y(t)( p q + q p) + X(t) q^{2} + {{Z(t) l^{2}}\over {q^{2}}}\bigr\rbrack \eqno (11).$$
The $(x- q)^{0}$ condition determines the time dependent global phase and normalization
included in $k$ which can be rearanged in the form $$ k(t) - k(0) =
i\int_0^{t}\Bigl(L(t') +  {{Z(t') l^{2}}\over {q^{2}}} - 2\hbar {\bigl(1
-\sqrt{(l/\hbar)^2 + 1/4 }\bigr)}\bigl(Z\beta - iY/2\bigr)\Bigr)dt', \eqno (12)$$ where
$L(t) = p(q,{\dot q}){\dot q} - H\bigl[ q, p(q,{\dot q}),t\bigr].$ Examining the three
terms in the expression (12) for $ k$, we see that the two first give
$i\bigl\{(pq) - p(0)q(0)\bigr\}/2$. The remaining terms in  $ k/\hbar$, namely  $ -
2i{\bigl(1 -\sqrt{(l/\hbar)^2 + 1/4 }\bigr)}\int_0^t \bigl(Z\beta - iY/2\bigr)dt'$ is
recognized as the angle (or phase) accumulated in the
nonadiabatic evolution, and contain a dynamical part $\theta_d (t)$  and a geometrical
one $\theta_H (t)$ (Hannay's angle).  

In the literature, $\theta_H (t)$ is usually
defined in relation with the introduction of time dependent canonical transformations.
However, a simple geometrical approach ( hence no calculatory) may be formulated to deduce
the decomposition of the total phase factor. Let us consider the classical equation
(9). The main property of this evolution is that is linear and area preserving. This
implies that any initial conditions $(Q(0), P(0))$ being at $t = 0$ on a centered ellipse
${\cal E}(0)$ in phase space evolve at time $t$ on a similar ellipse ${\cal E}(t)$ of
the same area. A little thought show that, more precisely, two points $M_0, N_0$ on 
${\cal E}(0)$ whose parameters differ by $\Delta\varphi$ evolve in points $M_t, N_t$ on 
${\cal E}(t)$ with the same difference of parameters so that the area 
$\overrightarrow{OM}_t\land \overrightarrow{ON}_t$ remains constant. The reason is that
the standard parameter $\varphi (\varphi \in [0, 2\pi])$ which parametrizes a point on
an ellipse is such that it is proportional to the area swept by the vector $\vec{v} =
\overrightarrow{OM}$. The natural origins of these familly (homothetical
centered) ellipses are the points associated with $\varphi = 0$, and the action 
angle coordinates $(I, \theta)$ of a point $M$ in phase space, with respect to the familly
associated with ${\cal E}(t)$, are respectively the area (divided by $2\pi$) and the usual
angular variable defined on this ellipse. There exist a natural transport (with respect
to the symplectic structure in phase space) for a familly of (homothetical centered)
ellipses. More precisely, let $M_t$ on ${\cal E}(t)$ a point of coordinates $(I, \theta
(t))$ and  ${\widetilde M}_{t+dt}$ its transported on ${\cal E}(t + dt)$. Clearly, this
transport must preserve areas, but this is not sufficient since it remains to precise how
one point (for example the origin) is transported: for this, one simply requires that the
area which is swept by the vectors $\vec{v}(\varphi)$ (on an ellipse) during the
transport has, when averaged over $\varphi$, a mean value equal to zero. Therefore the
transport is defined by $$\bigl<\overrightarrow{OM_t} \land \overrightarrow{M_t{\widetilde
M}_{t+dt}}\bigr> = 0.\eqno (13) $$ This transport associates to $M_t$ of coordinates 
$\theta (t),$ the point ${\widetilde M}_{t+dt}$ of coordinates  $\theta (t) + 
d\theta_H (t)\,\,(\theta_H (t)$: Hannay's angle). 

The angular coordinate on ${\cal E}(t + dt)$ of evolved point $ M_{t+dt}$ of $ M_t$ is 
$\theta (t) + d\theta (t)\, .$ The difference $d\theta (t) - d\theta_H (t) = d\theta_D (t)$
does not depend on the chosen point $ M_t$ on ${\cal E}(t )\, .$  The quantity 
$Id\theta_D = I(d\theta - d\theta_H)$ is the area  $\overrightarrow{O{\widetilde M}_{t+dt}}\land 
\overrightarrow{{\widetilde M}_{t+dt} M_{t+dt}}$ and can be written as 
the difference $\overrightarrow{OM_{t}}\land
\overrightarrow{M_{t}M_{t+dt}} - \overrightarrow{OM_{t}}\land
\overrightarrow{M_{t}{\widetilde M}_{t+dt}}$ of the area swept by the vector 
$\overrightarrow{OM_t}$ during its evolution and of the one swept by  
$\overrightarrow{OM}_t$ in the geometrical transport. Averaging these two areas on the
ellipse ${\cal E}(t)$, one can see that $Id\theta_D$ can be interpreted as the mean value
of the swept area during the motion. This justifies the appelation of dynamical angle for
$\theta_D (t)$.

To translate analyticaly the previous geometrical remarks, let $\vec E$ be a complex two
dimensional vector, it is known (for instance from optics) that one can describe
(homothetical centered) ellipses as the set of vectors $\overrightarrow{OM_t} = 
[{\vec E}(t) e^{-i\varphi}]$ and those obtained by the transport are represented 
by the set of vectors $\overrightarrow{O{\widetilde M}_{t+dt}} = [{\vec E}(t+dt) 
e^{-i d\theta_H}\, e^{-i \varphi}]$. Using the area preserving property
(proportional to $i{\vec E^{*}}\land{\vec E}$), one can easily verify that the
transport defined by Eq. (13) is also written as ${\vec E^{*}}\land (d{\vec E} -i 
d\theta_H {\vec E}) = 0$ or $${\dot\theta}_H = {{{\vec E^{*}}\land 
\vec{\dot E}}\over{i({\vec E^{*}}\land{\vec E})}}\, .\eqno (14)$$ Obviously Eq. (14)
gives new expression for the nonadiabatic Hannay angle of the generalized harmonic
oscillator. Within such formalism the above remarks justify that the general solution of
the Hamilton equation (9) may be looked for in the form $${\vec v}(t) =  
 A\,e^{-i(\theta (t) + \varphi)}{\vec E}(t),\ \ (\theta(0) = 0), \eqno
(15)$$ with $i{\vec E^{*}}\land{\vec E}$ conserved and $A$ and $\varphi$ fixed ($A$ and 
$\varphi$ are the conditions mesured with respect to the familly ${\vec E}(0)$). The
angular drift of the origins points of ${\vec E}(0)$ (measured with respect to ${\vec
E}(t)$) is naturally decomposed in a geometrical part (Hannay's angle) and a dynamical
one 
   $${\dot\theta} = {{{{\vec E^{*}}\land 
\vec{\dot E}}\over{i({\vec E^{*}}\land{\vec E})}}} + 
{{{\vec E^{*}}\land {\cal H}{\vec E}}\over{i({\vec E^{*}}
\land{\vec E})}}.\eqno (16)$$ (This relation is obtained by inserting ${\vec v}(t)$ in 
Hamilton equation (9) and making the wedge product with ${\vec v^{*}}(t)$).

If one wants to explicitly calculate $\dot{\theta}_H$ and $\dot\theta$, we must find
vector ${\vec E}(t)$. Before hand, one can impose the condition $-i({\vec E^{*}}\land{\vec
E}) = 4I$ (which fix the ellipse area)  and take the first componant of ${\vec E}$ to
be real. The general form of ${\vec E}(t)$ is thus $${\vec E}(t) = 
\pmatrix{{\sqrt{QQ^{*}}}\cr{{2i\beta (t)\sqrt{QQ^{*}} }}\cr}.\eqno (17)$$ It leads to the
relations $$\dot{\theta}_H = - i {{\dot\beta}\over{\beta + \beta^{*}}} - {i\over2}
{d\over{dt}}{\hbox{Ln}}(QQ^{*})\eqno (18) $$  
$$\dot\theta = - 2\bigl(Z\beta - iY/2\bigr)  - {i\over2}
{d\over{dt}}{\hbox{Ln}}(QQ^{*}).\eqno (19)$$ 

From the form of the time-dependent phase factors included in $k$ Eq. (12), it is clear
that $$\gamma_l (t) = - 2 {\bigl(1
-\sqrt{(l/\hbar)^2 + 1/4 }\bigr)}\int_0^{t}dt\bigl(Z\beta - iY/2\bigr)$$ $$ = 
{\bigl(1 -\sqrt{(l/\hbar)^2 + 1/4 }\bigr)}\Bigl[\Delta\theta (t) +
{i\over2} {\hbox{Ln}}\Bigl\{{{Q(t)Q^{*}(t)}\over{Q(0)Q^{*}(0)}}\Bigr\}\Bigr],\eqno (20)$$
we see that the logarithm term goes "downstairs" as time-dependent normalization factor in
the $\Psi_l(x,t) .$ The remaining term in $\gamma_l (t)$ is recognized as the phase factors
acquiered by the wave packet in its evolution.

Then, we can reach a simple relation between the geometrical phase for the quantum
singular oscillator and the nonadiabatic Hannay angle associated to the generalized
harmonic oscillator $$\gamma_l^{G} (t) = 
{\bigl(1 -\sqrt{(l/\hbar)^2 + 1/4 }\bigr)}\Delta\theta_H (t) ,\eqno (21)$$ where the
first part is independent of $\hbar$ and is equal to the Hannay angle, the second part
depends on $\hbar$ and $l$.

Now we turn our attention to the Eq. (3) which can be rewritten as a
simple Gaussian wave packet  $$\Psi_l(x,t) = \Bigl(x {{Q(0)Q^{*}(0)}\over{
Q(t)Q(t)}}\Bigr)^{\bigl(1/2 -\sqrt{(l/\hbar)^2 + 1/4 }\bigr)}\,
e^{-{1\over 4}{\hbox{Ln}}{{Q(0)Q^{*}(0)}\over{ Q(t)Q^{*}(t)}}}$$ $$
{\hbox{exp}}\Bigl({1\over\hbar}\Bigl\{-\beta x^{2}- 1/2 \bigl(l - ip(0)q(0)\bigr)
+ k(0)+ $$ $$\ \ \ \ \ \hbar
{\bigl(1 -\sqrt{(l/\hbar)^2 + 1/4 }\bigr)}\int_0^t\Bigl[  {{{{\vec E^{*}}\land 
\vec{\dot E}}\over{({\vec E^{*}}\land{\vec E})}}} + 
{{{\vec E^{*}}\land {\cal H}{\vec E}}\over{({\vec E^{*}}
\land{\vec E})}}\Bigr]dt'\Bigr\}\Bigr)\eqno
(22)$$ which is also a solution to the Schr\"odinger equation.

In conclusion, the squeezed Gaussian wave packet dynamics for the time-dependent singular
oscillator has been obtained as a simple squeezed wave packet. The quantum phases are
obtained explicitly and connected to the classical angle of the generalized harmonic
oscillator. The classical version of the generalized harmonic oscillator has been
discussed, and a new expression for the nonadiabatic Hannay angle has been obtained by
employing a geometrical approach of the evolution in phase space. In the literature,
$\theta_H (t)$ is usually defined in relation with the introduction of time dependent
canonical transformations. However, the geometrical part has been found by asking wether
there exists a natural transport (with respect to the symplectic structure) for a family
of (homothetical centered) ellipses. When the parameter $l$ vanishes, we see that the
Gaussian wave packet (22) corresponds to the evolution of the "ground" state of the
time-dependent generalized harmonic oscillator and the geometrical phase is equal to
one-half of the classical angle. This is just what was obtained in Ref. [7].\vfill \eject
\centerline {\bf References} \bigskip

\noindent [1] M.V. Berry, Proc. R. Soc. London Ser. A {\bf 392}, 45 (1984); J. Phys. A
{\bf 18}, 15 (1985).

\noindent [2] J.H. Hannay, J. Phys. A {\bf 18}, 221 (1985).

\noindent [3] Y. Aharonov and J. Anandan, Phys. Rev. Lett {\bf 58}, 1593 (1987).

\noindent [4] M.V. Berry and J.H. Hannay, J. Phys. A {\bf 21}, L325 (1988).

\noindent [5] M. Maamache, J.P Provost and G. Vall\'ee, J. Phys. A {\bf 23}, 5765 (1990),
Eur. J. Phys. {\bf 15}, 121 (1994).

\noindent [6] C. Jarzynski, Phys. Rev. Lett {\bf 74}, 1264 (1995).

\noindent [7] Y. C. Ge and M. S. Child, Phys. Rev. Lett {\bf 78}, 2507 (1997).

\noindent [8] J. Lie, B. Hu and B. Li,  Phys. Rev. Lett {\bf 81}, 1749 (1998).

\noindent [9] M. Maamache, J.P Provost and G. Vall\'ee, Phys. Rev. A {\bf
59}, 1777 (1999).

\noindent [10] D. Y. Song, Phys. Rev. Lett {\bf 85}, 1141 (2000).

\noindent [11] E. J. Heller, J. Chem. Phys. {\bf 65}, 4979 (1976).

\noindent [12] B. Sutherland, J. Math. Phys. {\bf 12}, 246 (1971).

\noindent [13] F. Calogero, J. Math. Phys. {\bf 12}, 419 (1971).

\noindent [14] S.M Chumakov, V.V Dodonov and V.I Man'ko, J. Phys. A {\bf 19}, 3229 (1986).

\noindent [15] V.V Dodonov, V.I Man'ko and D.V Zhivotchenko, IL Nuovo Cimento B {\bf 108},
1349 (1993).

\noindent [16] M. Maamache,  Phys. Rev. A {\bf 52}, 936 (1995); A {\bf 61}, 026102 (2000);
J. Phys. A {\bf 29}, 2833  (1996);  Ann. Phys. (N.Y.) {\bf 254}, 1 (1997).

\noindent [17] R. S. Kaushal and D. Parashar, Phys. Rev. A {\bf 55}, 2610 (1997).

\noindent [18] I.A. Pedrosa, G.P. Serra and I. Guedes, Phys. Rev. A {\bf 56}, 4300
(1997). 

\noindent [19] V.V Dodonov, V.I Man'ko and L. Rosa, Phys. Rev. A {\bf 57}, 2851
(1998).

\end